# Controlled generation, manipulation and frequency conversion of broadband orbital angular momentum spectrum of asymmetric optical vortex beams

Sabir Ul Alam,† A. Srinivasa Rao,† Anirban Ghosh, Pravin Vaity,** and G. K. Samanta*

Photonic Sciences Lab., Physical Research Laboratory, Navarangpura, Ahmedabad 380009, Gujarat, India
*Corresponding author: gsamanta@prl.res.in
†Authors having equal contributions. ** Present address: Laboratory for Microactuators, IMTEK, University of Freiburg, Germany.

**We report on generation and control of tunable, broad orbital angular momentum (OAM) spectrum of a vortex beam. Using two spiral phase plates (SPPs), we have converted the Gaussian beam of Yb-fiber femtosecond laser at 1064 nm into optical vortices of orders, $l$=1-6, with conversion efficiency >95%. By adjusting the transverse shift of the SPPs with respect to the incident Gaussian beam axis, we have transformed the symmetric (intensity distribution) optical vortex of order, $l$, into an asymmetric vortex beam of broad OAM spectrum of orders, $l$, $l$-1, $l$-2 ...0. While the position of the SPPs determines the weightage of the OAM modes, however, the weightage of the higher OAM modes transfer to lower OAM modes with the shift of SPPs and finally resulting a Gaussian beam ($l$=0). Using single pass frequency doubling of asymmetric vortices in a 5-mm-long bismuth triborate crystal, we have transferred the pump OAM spectra, $l$, $l$-1, $l$-2...0, into the broad spectra of higher order OAM modes, 2$l$, 2$l$-1, 2$l$-2...0 at green wavelength, owing to OAM conservation in nonlinear processes. We have also observed that the conversion efficiency of the frequency-doubled vortices of all orders increases with the increase in the asymmetry of the pump vortices. Using pump power of 4.6 W, we have generated symmetric vortices at green wavelength of output powers, 1.2, 0.76, 0.47, 0.4, 0.32, and 0.3 W and orders of $l_s$=2, 4, 6, 8, 10, and 12, respectively.**

Optical vortices, the higher order Laguerre-Gaussian (LG) beams of zero radial index, have screw-like (helical) phase dislocation in their wavefront. The existence of helical phase variation along propagation, results the beams to carry phase singularity [1]. The doughnut shaped intensity distribution of such beams appear due to the undefined phase at the singular point and corresponding vanishing intensity. The phase of the optical vortices around the singular point varies from 0 to $2\pi l$, where, $l$, is the topological charge/order of the optical vortex. Like the circularly polarized light beams having spin angular momentum (SAM) of $\pm\hbar$ per photon, the optical vortex beams of orders $\pm l$ carry orbital angular momentum (OAM) of $\pm l\hbar$ per photon [2]. The sign ($\pm$) indicates the polarizations (left and right circular), and the direction of phase variation (clockwise and counter-clockwise). Unlike SAM having two allowed states (+1 and -1), the OAM has a discrete spectrum defined over an infinite-dimensional Hilbert space. Therefore, the OAM can be used for super dense coding and quantum communication. In fact, the doughnut intensity distribution and the existence of OAM make the optical vortices indispensable for the variety of applications in science and technology including classical and quantum communications [3, 4], particle micromanipulation and lithography [5], high-resolution microscopy [6], interferometry [7], and material processing [8].

Optical vortices are generally produced through the mode conversion of laser beams of Gaussian spatial profile using standard mode converters such as cylindrical lenses [9], spiral phase plates (SPPs) [10], q-plates [11] and dynamic phase modulation through the computer generated holography technique based spatial light modulators (SLMs) [12]. However, all these mode converters, while designed to produce a vortex beam of particular order, $l$, result the output beam having pure OAM mode of order, $l$. In addition, these mode converters suffer from at least one of the common drawbacks such as, limited wavelength coverage, low power handling capabilities, and high costs. Over the decades a variety of new techniques [13-15] have been proposed and tested to produce high power and higher order vortex beams at different wavelengths. On the other hand, efforts have been made to produce broad OAM states in a single beam through the superposition of vortex beams of different OAM states [16, 17]. However, the realization of next generation envisaged projects towards classical and quantum communication [3, 4] and new experiments in the fields of classical and quantum optics demand optical vortex beams of wide OAM spectra extended over wide wavelength ranges. Here, we report on the experimental generation of optical vortex beam of tunable broad OAM spectra in a single beam by incorporating the asymmetry in the intensity distribution of the beam. We have also transferred the OAM spectra of the pump to a broad spectra of higher order OAM modes to new wavelength through nonlinear frequency conversion process. To the best of our knowledge, this is the first experimental report on the measurement of OAM content of asymmetric vortex beams of different orders and their frequency doubling characteristics.

Schematic of the experimental setup is shown in Fig. 1. A 5 W Yb-fiber laser (Fianium, FP1060-5-fs) delivering ultrafast pulses of temporal width (full width at half maximum, FWHM) of 260 fs in $sech^2$ shape at a repetition rate of 78 MHz is used as the fundamental (pump) laser source. The output of laser has a Gaussian intensity profile and a spectral width of 15 nm centred around 1060 nm. A motorized power attenuator comprised with a pair of reflecting polarizers and a half wave plate, $\lambda/2$, is used to control the average power of the pump laser. Using the combination of the spiral phase plates, SPP1 and SPP2, having phase-winding corresponding to vortex ($LG_0^l$, radial and azimuthal indices are zero and $l$) beam of orders, $l = 1$ and 2, respectively, and the vortex doubler [18] setup comprised with a polarizing beam splitter cube (PBS), quarter wave plate, $\lambda/4$ and a mirror, M, we can generate optical vortex beams of orders, $l = 1$ to 6. The working principle of the vortex doubler can be found elsewhere [18]. The SPPs produce symmetric vortex beam when the input Gaussian beam having amplitude distribution symmetric about the propagation axis pass through the axis of the SPPs at normal incidence. However, any misalignment, in terms of tilt and or displacement between the beam propagation axis and the axis of the SPPs, creates asymmetry in the intensity distribution of the generated vortex beam.

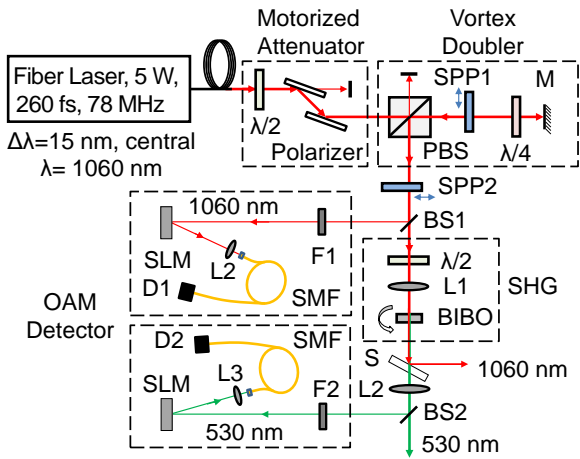

Fig. 1. Experimental setup for asymmetric vortices. $\lambda/2$, half-wave plate; PBS, polarized beam splitter cube; SPP1,2, spiral phase plates; $\lambda/4$, quarter-wave plate; BS1,2, beam splitters; BIBO, nonlinear crystal; S, wavelength separator.; F1,2, filters; L1-4, lenses; SLM, spatial light modulators; SMF, single mode fibre, D1-2, detectors.

To create asymmetric vortex beams we have mounted SPPs on a $xy$-translation stage with least count of 10 μm. A fractional part of the pump beam is extracted with the help of the beam splitter (BS1) and analysed with the OAM detection unit consists of a spatial light modulator, SLM (Hamamatsu, X13138-08) and a single mode fibre coupled power meter. The lens (L1) of focal length, $f = 750$ mm is used to focus the pump beam at the centre of the nonlinear crystal. A 5 mm long and 4 x 8 mm² in aperture bismuth triborate (BIBO) crystal cut for type-I ($e + e \rightarrow o$) SHG in optical $yz$-plane ($\varphi = 90°$) with an internal angle of $\theta = 168.5°$ at normal incidence is used for collinear frequency doubling of 1064 nm. The crystal faces are anti-reflection coated for both fundamental and second-harmonic (SH) wavelengths. The second $\lambda/2$ is used to adjust the polarization of the pump beam with respect to the crystal orientation for phase matching. The frequency doubled green is extracted from the pump beam using a wavelength separator (S) and subsequently collimated with the help of a lens (L2) of focal length, $f = 300$ mm. The OAM of the frequency doubled asymmetric vortex in the green is analysed with the help of the beam splitter (BS2), and the OAM detection unit consists of SLM (Holoeye, LC-R-2500) and single mode fibre at green. Filters, F1 and F2 are used to remove any unwanted radiation from the pump and the green beams, respectively.

To study OAM content of the vortex beams with the asymmetry in the dark core position, we have generated pump vortex beam of order, $l_p = 3$, using SPP1 and SPP2 in tandem. We have created the asymmetric position of the dark core by moving the SPPs along the vertical direction with a normalized distance, $x_o/w_p$ ($w_p = 1.5$ mm, is the beam waist radius of the Gaussian beam at the front surface of the SPP) away from the beam axis. The results are shown in Fig. 2. As evident from the first column, (a-d), of Fig. 2, the vortex beam of order, $l_p = 3$, has a symmetric doughnut intensity distribution (Fig. 2a) for, $x_o/w_p = 0$, with dark core at the centre, however, the dark core moves away from the centre of the beam with the transverse shift of, $x_o/w_p = 0.13, 0.4$ and 0.67 of the SPPs producing asymmetric intensity distribution of the vortex beam. Although, we have restricted the shift of the SPP up to a normalized distance of, $x_o/w_p = 0.67$, however, for $x_o/w_p \gg 1$ one can expect the resultant beam to be a Gaussian beam ($l = 0$) as predicted in Ref. [19]. We also analysed the azimuthal phase variation of the asymmetric vortex ($LG_0^l$) beams using a balanced polarization Mach-Zehnder interferometer (MZI) [20] with a Dove prism in one of its arms. The extra reflection due to the Dove prism produces a superposition state, ($LG_0^l + LG_0^{-l}$), at the output of the MZI and subsequent diagonal projection of this state produces ring lattice structure with $2l$ numbers of radial fringes or petals [21] for the input beam of vortex order, $l$. The second column, (e-h) of Fig. 2 show the ring lattice structure of six petals (as expected) with equal intensity for the symmetric vortex beam of order, $l_p = 3$, however, some of the petals gradually loses their intensity with the asymmetry to the vortex beam. Since the ring lattice structure maintains its shape, a signature of the azimuthal phase variation of the vortex beams corresponding to its intrinsic topological charge, even for the shift of the SPP to a normalized distance of, $x_o/w_p = 0.67$, it is difficult to predict the order or the OAM content of the asymmetric vortices using polarization interferometry and projective measurement techniques as used for fractional vortices [22]. While, the loss in the intensity of the petals can be attributed to the asymmetric intensity distribution of the vortex beam, the appearance of lattice structure confirms the presence of OAM modes in the asymmetric vortices.

We have measured the OAM content of the asymmetric vortex beam using the OAM detector based on mode-projection technique [4], where the vortex beam is diffracted by the SLM having blazed fork grating of different vortex orders. The first order diffracted beam is coupled to a single mode fibre and subsequently measured using a power meter. As evident from the third column, (i-l), of Fig. 2, the symmetric vortex ($x_o/w_p = 0$) beam produced by the combination of SPP1 and SPP2 has pure OAM mode of order, $l_p = 3$, with negligible contribution from other OAM modes. However, with the increase in the asymmetry, $x_o/w_p = 0.13, 0.4$ and 0.67, we observe the pure OAM mode is transforming into a mixture of lower order OAM modes of different weightage. The asymmetry in the vortex beam of order, $l$, produces a broad OAM spectra in the beam with orders, $l$, $l$-1, $l$-2 ...0. The weightage of the OAM modes is transferred from the higher to lower OAM modes with the increase in vortex asymmetry and finally producing an OAM mode of order, $l = 0$ (Gaussian beam) at very high asymmetry, $x_o/w_p \gg 1$. Unlike Ref. [19], here, we observed the integer OAM modes in the off-axis vortices. From this study it is evident that one can tailor the OAM spectra of a pure vortex beam by simply adjusting the asymmetry in the beam. Therefore, we can conclude that by using a single SPP of any order, one can control the OAM spectra of the vortex beam to produce pure OAM mode as well as a precise mixture of OAM modes by simply shifting the axis of the SPP away from the beam axis.

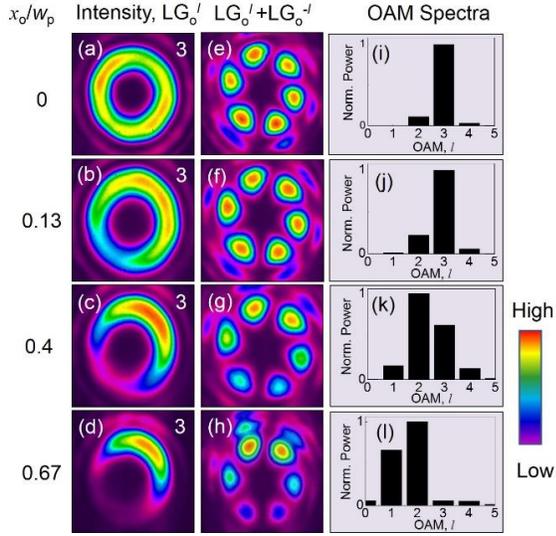

Fig. 2. Variation of far-field intensity distribution, (a-d), corresponding characteristic lobes, (e-h), and the measured OAM spectra, (i-l), of pump vortex beam of order, $l_p$=3, at different values of transverse shift ($x_o/w_p$) of the SPPs away from the beam axis.

Knowing the mixed OAM content of the asymmetric vortices, we have studied the transfer of OAM modes of the asymmetric vortices in nonlinear frequency conversion process. Using the pump vortices of orders, $l_p = 3$ and 2, and shifting the SPPs we have recorded the intensity profile, interference pattern and the OAM content of the SHG beam with the results shown in Fig. 3. As evident from the first column, (a-c), of Fig. 3, the SHG beam has a symmetric intensity distribution for the symmetric pump vortex ($l_p = 3$) beam ($x_o/w_p = 0$), however, with change in the value of, $x_o/w_p = 0.4$ and 0.67, the SHG beam has higher asymmetry in its intensity distribution as compared to that of the pump vortices. Such observation can be attributed to the fact that the nonlinear frequency conversion processes are intensity dependent phenomenon and the lower intensity part of the asymmetric vortex beam has lower SHG conversion than that of the higher intensity part of the beam resulting a highly asymmetric SHG beam. To observe the effect of azimuthal phase variation of the asymmetric vortices generated in SHG process, we have recorded the interference pattern of the SHG beam using the polarization MZI. The second column, (d-f), of Fig. 3, shows the ring lattice structure with 12 number of petals for the SHG beam of the symmetric pump vortex of order, $l_p = 3$, confirming the order of the SH vortex to be $l_{sh} = 6$. However, it is interesting to note that the highly asymmetric SHG vortex beams also maintain the lattice structure, a signature of azimuthal phase variation, with lower number of lobes. We have measured the OAM modes of the SHG beam as shown in third column, (g-i) of Fig. 3. Similarly, using pump vortex beam of order, $l_p = 2$, and varying the asymmetry with different values of, $x_o/w_p = 0$, 0.4 and 0.67, we also observed the variation in the intensity profile, ring lattice structure, and the OAM content with the results shown by fourth, (j-l), fifth, (m-o), and sixth, (p-r), columns of Fig. 3, respectively. As expected, the symmetric vortex beam of order, $l_p = 3$, and 2, produces SHG vortex beams of order, $l_s = 2 \times l_p = 6$ and 4, respectively, owing to OAM conservation in SHG process. The broad OAM spectra of orders, $l$, $l$-1, $l$-2 ...0 of the asymmetric pump vortex is transferred to the SHG beam producing broad OAM spectra at new wavelength with orders, $2l$, $2l$-1, $2l$-2 ...0. The weightage of the OAM modes of the SHG beam is determined by the weightage of the OAM modes of the pump beam. Such observation clearly shows the possibility of generation of broad OAM spectra at different wavelengths across the electromagnetic spectrum by simply adjusting the asymmetry in the pump vortex beam of the nonlinear frequency conversion processes. It is interesting to note from the results of Fig. 3 that the rate of transfer of the weightage of high order OAM modes to the lower order modes and finally to Gaussian mode is higher for lower order vortices.

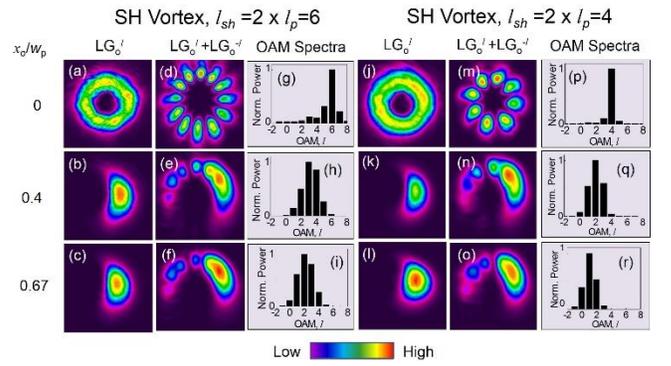

Fig. 3. Far-field intensity distribution, (a-c), interference pattern, (d-f), and the variation OAM spectra, (g-i), of the SHG beam of pump order, $l_p = 3$, and the far-field intensity distribution, (j-l), interference pattern, (m-o), and the variation of OAM spectra, (p-r), of the SHG beam of the SHG beam of pump order, $l_p = 2$, for different asymmetries in the pump vortices.

We also studied the single-pass conversion efficiency of the vortex beams with different asymmetry. Using a lens of focal length, $f = 100$ mm, we have focused the pump vortex beams of power, 2.5 W and vortex orders, $l_p = 1$, 3 and 5, at the centre of the nonlinear crystal and measured the power and single-pass efficiency of the frequency-doubled vortex beams for different asymmetry in the pump vortex beams. The results are shown in Fig. 4 (a). As evident from Fig. 4(a), the symmetric vortex beam of orders, $l_p = 1$, 3 and 5, produce symmetric SHG vortex beams of orders, $l_{sh} = 2 \times l_p = 2$, 6 and 10, and output powers of 0.6, 0.18 and 0.16 W at a single-pass conversion efficiency of 24%, 7% and 6.7%, respectively. The decrease in the conversion efficiency with the vortex orders can be attributed to the increase of the dark core size of the vortices with its orders [13,18]. However, it is interesting to note that the conversion efficiency and the SHG power increase with the increase in the asymmetry of the pump vortices. For pump vortex of order, $l_p = 1$, the single-pass conversion efficiency (power) increases from 24% (0.6 W) to 32% (0.8 W) with the increase in the asymmetry of the pump vortex for the increase of $x_o/w_p$ from 0 to 0.8. However, for further increase of $x_o/w_p$ up to 1.3 results SHG conversion efficiency constant at ~33%. On the other hand, for the pump vortices of orders, $l_p = 3$ and 5, the SHG efficiency increases from 7% to 23.5% and 6.7% to 14% for the increase of $x_o/w_p$ from 0 to 1.3, without any sign of saturation. To get better insight on the increase of SHG efficiency with asymmetry in the pump vortices of all orders, we have recorded the intensity profile of the pump vortices with different asymmetries. The results are shown in Fig. 4(b). As evident from Fig. 4(b), the lower order vortex loses its symmetric intensity pattern faster with the increase in the value of $x_o/w_p$ and produce intensity pattern resembling Gaussian beam. Therefore, for pump vortex beam of order, $l_p$ = 1, (see first row of Fig. 4(b)), the SHG efficiency increases with the increase of $x_o/w_p$ from 0 to 0.8 as the entire beam power is accumulated at the side opposite to the shift direction of the SPP. However, with further increase in $x_o/w_p$, does not change the beam shape and the pump intensity, resulting constant SHG conversion. On the other hand, as evident from the intensity patterns of second and third rows of Fig. 4(b), corresponding to the pump vortices of orders $l_p$ = 3 and $l_p$ = 5, respectively, the beam intensity at one side of the beam is gradually increasing with the increase of $x_o/w_p$ from 0 to 1.3 and subsequently increasing the SHG efficiency. From this study it is evident that by incorporating asymmetry in the vortex beam one can generate mixed OAM modes at different wavelengths at higher conversion efficiency than the pure OAM mode. Using same focusing condition and pump power of 4.6 W, we have generated symmetric SHG vortices of output power of 1.2, 0.76, 0.47, 0.4, 0.32, and 0.3 W and orders of, $l_{sh}$ = 2, 4, 6, 8, 10, and 12 respectively. Unlike our previous report [18], here, we have generated SHG vortices of higher power due to the use of longer length of the BIBO crystal. We have also studied the spectral and temporal width of the asymmetric vortex beams using a spectrometer and intensity autocorrelator. As expected, we did not observe any change in the spectral and temporal width of the asymmetric vortices of all orders. The typical spectral and temporal width of the green vortices are measured to be ~2.4 nm centred at 532 nm and ~200 fs, respectively [18]. Like Ref. [23], we also observed the rotation of the asymmetric vortices of all orders due to the Gouy phase at both fundamental and second harmonic wavelengths.

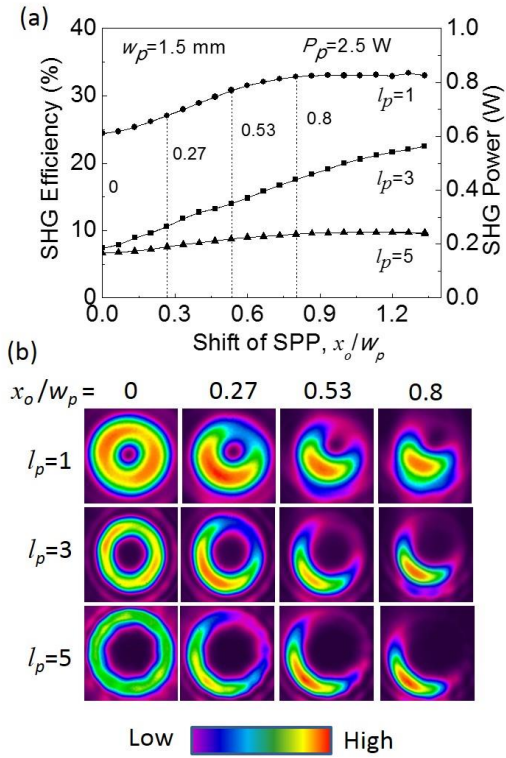

Fig. 4. (a) Variation of conversion efficiency and power of the frequency doubled asymmetric vortex beams for different values of transverse shift, ($x_o/w_p$), of SPP in the pump beam. (b) The variation of the intensity distribution of pump vortices of orders, $l_p$ = 1, 3, and 5, for different values of $x_o/w_p$.

In conclusion, we have demonstrated the transformation of pure OAM mode into a mixture of OAM modes in single beam by simply shifting the axis of the SPPs with respect to incident Gaussian beam. Using two SPPs we have generated pure OAM modes of order as high as $l$ = 6 and produced tunable OAM spectra with orders, $l$, $l$-1, $l$-2,...0, at different weightage factors. We have observed that the increase in the asymmetry of the vortex beam, transfers the OAM modes from higher orders to lower orders and finally producing Gaussian beam ($l$=0). Using nonlinear frequency conversion we have observed that the broad spectra of the pump OAM modes can be transferred into new wavelength to produce broad OAM spectra with higher order OAM modes owing to the OAM conservation in nonlinear optical processes. We have also observed that with the transfer from the symmetric to asymmetric intensity distribution of the vortex beam, the single-pass SHG efficiency increases due to the transition of the intensity distribution of the beam from annular shape to the Gaussian shape. This generic approach can be used to produce broad OAM spectrum using higher order SPPs and at different wavelengths with the help of different nonlinear interactions.